\documentclass[traditabstract]{aa} 
\usepackage{graphicx}
\usepackage{txfonts}
\usepackage{natbib}
\def\lref{\hbox{$\lambda_\mathrm{ref}$\,}}
\begin{document}
  \title{Synthesising, using, and correcting for telluric features in high-resolution astronomical spectra}
    \subtitle{A near-infrared case study using CRIRES\thanks{Based on observations made with ESO Telescopes 
              at the Paranal Observatory under programme IDs 60.A-9076(A), 079.D-0810(A), and 179.C-0151(A)}}

   \titlerunning{Synthesising and correcting telluric features}

   \author{A. Seifahrt\inst{1,2,3}
          \and
           H.~U. K\"{a}ufl\inst{3}
          \and
           G. Z\"{a}ngl\inst{4,5}
          \and
           J.~L. Bean\inst{2}
          \and
          M.~J. Richter.\inst{1}
          \and
           R. Siebenmorgen\inst{3}
 }
   
     \institute{Physics Department, University of California, Davis, CA 95616, USA\\
\email{seifahrt@physics.ucdavis.edu}
\and
Universit\"at G\"ottingen, Institut f\"ur Astrophysik, Friedrich-Hund-Platz 1, D-37077 G\"ottingen, Germany
     \and
     European Southern Observatory (ESO), Karl-Schwarzschild-Str. 2, D-85748 Garching, Germany
     \and
     Ludwig-Maximilians-Universit\"at M\"unchen, Meteorologisches Institut, Theresienstrasse 37,  80333 Munich, Germany
 \and
     Deutscher Wetterdienst, Frankfurter Str. 135, 63067 Offenbach, Germany
      }

   \date{as of \today}

 
  \abstract
{We present a technique to synthesise telluric absorption and emission features both for in-situ 
wavelength calibration and for their removal from astronomical spectra. While the presented technique is applicable for
a wide variety of optical and infrared spectra, we concentrate in this paper on selected high-resolution near-infrared spectra
obtained with the CRIRES spectrograph to demonstrate its performance and limitation. We find that synthetic spectra reproduce telluric
absorption features to about 2\%, even close to saturated line cores. Thus, synthetic telluric spectra could be used to replace
the observation of telluric standard stars, saving valuable observing time. This technique also provides a precise in-situ
wavelength calibration, especially useful for high-resolution near-infrared spectra in the absence of other calibration
sources.}


   \maketitle
%

\section{Introduction}
Ground based astronomical observations suffer from the disturbing effects of the Earth's atmosphere. Strong ro-vibrational line systems of water 
vapour (H$_2$O), carbon dioxide (CO$_2$), and ozone (O$_3$) shape the observational windows in the near infrared that define the bandpasses 
of the $J$, $H$, $K$, $L$, and $M$ filters from 900 to 5500~nm. In addition, nitrous oxide (N$_2$O) and methane (CH$_4$) contribute noticeably 
to the atmospheric transmission losses in the near infrared, hampering the observation of important astrophysical lines.

The Earth's atmosphere below about 60\,km is in local thermodynamic equilibrium and Kirchhoff's law applies. While the transitions of the molecules
discussed above appear in absorption in a stellar spectrum, they also appear in emission on the sky since the Earth's atmosphere is radiating its 
thermal energy into space in these lines. The emission originating from the Earth's atmosphere competes with the astronomical signal at wavelengths 
longer than about 2300~nm, depending on ambient temperature, slit width, and the readout noise of the detector. Because their spatial signature is 
extended and moderately flat, beam-switching (nodding) techniques or in-situ fitting and subtraction in 2D spectra is generally sufficient to remove 
the direct influence of the emission lines in the near and mid-infrared.

The removal of the absorption features is considerably more difficult because no in-situ reference for these features is available in the optical and 
near-infrared. The usual technique is to observe a star with either very few or well determined intrinsic features (a ``telluric standard'') close in time 
and airmass to the scientific target. The choices for such a standard are early-type stars (spectral types mid B to late A) that exhibit a rather clean 
continuum despite strong hydrogen lines, or solar-like stars of spectral type early to mid G. In the latter case, high resolution Fourier transform spectra 
\citep[e.g.,][]{ACE,fts} of the sun provide the necessary template to correct for the intrinsic stellar lines of the standard star.

Despite the wide use of this empirical calibration \citep[e.g.,][and references therein]{Vacca03}, the use of standard stars for removing telluric absorption 
lines has several disadvantages, especially when the correction is needed for low resolution spectra. At a spectral resolving powers exceeding 
$R\sim20\,000$, telluric lines start becoming resolved and bandpasses with high telluric line density become accessible to ground based observations.
The full potential of high resolution spectroscopy can only be used when unsaturated telluric lines are removed without compromising the 
intrinsic signal-to-noise ratio (SNR) of the data. To reach noise limited performance, the path length (airmass) and instrumental profile between the 
standard star and the science target must match. Also, the standard must be bright enough not to limit the final SNR. 

The intrinsic lines of early type standard stars, even when being accounted for, make the precise study of near-infrared hydrogen lines very difficult \citep{Vacca03}
and the choice of a fast rotator can be an advantage. Also, the mismatches in the line depths of solar-type standard stars with
the solar FTS atlas, caused, e.g., by differences in metallicity or effective temperature, hamper precise abundance studies from near-infrared metal lines.
 \begin{table*}[th]
      \caption[]{Log of CRIRES observations retrieved from the ESO archive.}\label{tab:obslog}
\centering 
\begin{tabular}{c c c c c c c c c } 
\hline\hline 
Object & Spectral- & Date \& Time & \lref & $v$ & Slitwidth & $R^\mathrm{\,b}$ & Air- & SNR$^\mathrm{c}$ \\ 
       & type$^\mathrm{a}$ &      & (nm)  & (GHz) & (\arcsec) & & mass &      \\
\hline 
HD 74575  & B1.5III& 2007-05-10 01:00 UT & 1134.0 & 264,367 & 0.2 & 64,000 & 1.30 & 250\\
HD 148605 & B3V & 2007-03-03 09:30 UT & 1504.8 & 199,224 & 0.2 & 80,000 & 1.03 & 200\\ 
HD 71155  & A0V & 2007-02-28 04:30 UT & 2076.5 & 144,374 & 0.4 & 60,000 & 1.24 & 300\\
HD 121847 & B8V & 2008-02-29 08:30 UT & 2336.2 & 128,325 & 0.4 & 65,000 & 1.01 & 300\\
HD 118716 & B1III&2007-03-05 05:00 UT & 3907.8  & 76,716 & 0.2 & 65,000 & 1.30 & 380\\
HD 177756 & B9Vn& 2007-04-22 10:00 UT & 4760.8 & 62,971 & 0.2 & 87,000 & 1.06 & 200\\
\hline 
\end{tabular}
\begin{list}{}{}
\item[$^{\mathrm{a}}$] SIMBAD spectral type
\item[$^{\mathrm{b}}$] Estimated resolving power from fitting telluric lines in the analysed spectral region.
\item[$^{\mathrm{c}}$] Estimated signal-to-noise ratio in the continuum of the analysed spectral region. 
\end{list}
\end{table*}
Another problem arises for the $L$ and $M$ band, where the source flux of early- and solar-type standard 
stars can be the limiting factor, favoring the observation of non-stellar sources, like bright asteroids. 

Additional limitations in the use of empirical telluric corrections may be based on the spectrograph itself. Limited spectral coverage around the hydrogen lines of early-type
standard stars prohibit fitting and correcting these lines, which leaves a disturbed continuum in the spectrum of the science target. Adaptive optics fed long-slit spectrographs, 
such as CRIRES at the VLT (see Sect.~\ref{sec:Observations}), introduce another level of complication. The diffraction-limited full width at half maximum (FWHM) of the stellar 
point spread function (PSF) can be considerably smaller than the slit width, so that one often observes with an under-filled entrance slit. The actual instrumental profile, and hence 
spectral resolving power\footnote{Throughout this paper we define the spectral resolving power as $R = \lambda/\Delta\lambda$, where $\Delta\lambda$ is the FWHM of the 
instrumental profile.}, is then governed by the PSF of the source, rather than by the slit width of the spectrograph. Moreover, slight displacements of the source in the slit will lead 
to shifts in the wavelength at the detector plane. In the case of CRIRES, a shift of only 30~m/s (1/50 of a CRIRES pixel or 4~mas at the entrance slit) can notably degrade the 
telluric correction (see, e.g., Sec.~\ref{CO2}). Since most of the telluric lines in the near-infrared are not fully resolved, even at $R$=100\,000, their shapes are governed by the 
instrumental profile. 
As a consequence, even at the same airmass the telluric standard star cannot perfectly correct the telluric features in the spectrum of the science target because differences in 
the performance of the adaptive optics ultimately lead to differences in the instrumental profile and variations of the wavelength solution. A further drawback in the use of telluric 
standard stars is the waste of observing time for the acquisition and observation of these stars, especially under the constraint of matching airmasses at small time differences.
As mentioned earlier the problem of source brightness becomes especially apparent at longer wavelengths where one can easily spend as much or even more time observing a 
telluric standard than a science target. 

While telluric emission and absorption features have been discussed so far as a disturbance in astronomical spectroscopy, they can as well turn out to be very useful. 
Determining the wavelength solution of near infrared spectra, especially at high spatial and spectral resolution, proves to be very difficult. Rare gas emission line lamps used in low 
resolution near-infrared spectrographs, such as He, Ne, Xe, and Kr lamps, provide much too sparse of a line coverage to be useful at high spectral resolution. Even the rich
spectra of ThAr emission line lamps, commonly employed by high resolution optical spectrographs, have a much lower line density in the near-infrared than in the red optical. 
Typical line densities of $\sim400$ lines per $\Delta\lambda=100$\,nm around $\lambda=1000$~nm drop quickly to less than $\sim20$ lines per $\Delta\lambda=100$\,nm at $\lambda=2500$~nm 
\citep{Kerber08}. Given the typical wavelength coverage of $\sim\lambda/200$ per detector, many CRIRES spectral settings remain poorly calibrated. Moreover the dynamic
range of the ThAr emission spectrum is very high, owing to the strong contrast between the Th and Ar lines, leaving weaker lines at low SNR while nearby stronger lines saturate 
quickly. 

Recently frequency combs \citep{comb} have been proposed for the ideal spectral calibration of high resolution spectrographs. In principle a frequency comb would provide for high line 
density while linking the dispersion solution of the spectrograph to the time standard. However, the line density of a simple comb is much too high, so that some optical filtering is required, 
which will produce detrimental artifacts. 
Also, a frequency comb might solve for the problem of insufficient line density and uniformity of emission line lamps but the instrumental profile will still be different than for the science target,
ultimately limiting the final precision of the wavelength calibration of the science spectrum. 

This limitation applies also for atmospheric emission lines from ro-vibrational levels of the Hydroxyl radical OH$^*$ (Meinel bands, commonly referred to as OH$^*$ airglow emission) 
which are routinely used for the calibration of low-resolution near-infrared spectrographs \citep[e.g.,][]{Rousselot00}. Moreover, neither the line strength nor the line density is sufficient 
to be used as a useful wavelength calibration over a wide spectral range at high spectral and spatial resolution, as provided by CRIRES. The small spatial pixel scale of CRIRES (86mas/pix) 
makes the use of OH$^*$ lines for calibration purposes challenging, given that each pixel is less than $10^{-2}$ arcsec$^2$ on the sky. In contrast, the atmospheric absorption and 
emission features of other molecules show a high density of strong absorption lines and thus provide a natural in-situ wavelength calibration. Moreover, the telluric absorption lines suffer 
the same instrumental effects as the lines in the stellar spectra and, hence, provide an intrinsically more precise calibration source as compared to telluric emission lines or lamp emission lines, 
the latter even recorded in separate exposures and often hours apart from the science spectra.

We have thus investigated the feasibility of calculating theoretical atmospheric transmission spectra in order to use them for wavelength calibration and to remove their signature from 
the astronomical spectra, thus, replacing standard stars as a source of telluric calibrators. This idea is not new and has been used successfully before \citep{Lallement93,Widemann94,Bailey07}. 
Nevertheless, modeling telluric spectra has not evolved to a standard technique in optical and infrared spectral data reduction because suitable radiative transfer codes were hard to 
access and synthesising adequate spectra was hindered by the incompleteness of molecular line databases. This situation has improved in the last years.

In an effort to characterise the abilities and the efficient usage of the CRIRES spectrograph, we have developed a general method to model the transmission and emission
spectrum of the Earth's atmosphere above Cerro Paranal. This method can be used to perform a wavelength calibration of CRIRES and to subsequently remove the telluric absorption 
lines apparent in these spectra. Moreover the model can also be used to predict the performance of planned CRIRES observations using a forecast of the atmospheric conditions during a 
planned observing run. 

The paper is structured as follows: In Sect.~\ref{sec:Observations} we describe the data reduction of the CRIRES observations. In Sect.~\ref{sec:Synthesis} we describe the the radiative transfer code
and input parameters used to synthesise the telluric spectra, which we compare to the observations in Sect.~\ref{sec:performance}. We summarise our findings in Sect.~\ref{sec:Summary} and close
with conclusions about the performance of the presented technique.

\section{Observations and data reduction} \label{sec:Observations}

CRIRES is a high resolution ($R\leq100\,000$) near-infrared ($\lambda\lambda$=960--5200\,nm) adaptive optics (AO) - fed spectrograph at the VLT on Paranal, Chile 
\citep{kaufl04,kaufl06a,kaufl06b,jerome06}.  Commissioning and science verification (SV) observations were executed in October 2006 and February 2007. 
The spectrograph is available to the community and has been used for regular observations since April 2007. 

We selected observations from both SV periods as well as more recent observations from the ESO data archive to test the telluric model spectra described in Sect.~\ref{sec:Synthesis}. For this
purpose we chose observations of standard stars at different wavelengths, observed at different airmasses with nominal spectral resolving powers of $R=50\,000$ and 100\,000. 
The log of observations is given in Tab.~\ref{tab:obslog}.

Observations were always obtained in an AB or ABBA nodding pattern. Data reduction followed the standard steps for long-slit spectrographs as applicable to the CRIRES data format. 
All raw frames were treated with a non-linearity correction before pairwise subtraction removed the atmospheric emission features. The individual A-B and B-A 
frames were then divided by a normalised flatfield. Optimally extracted 1D spectra at both nodding positions were obtained by using a custom made IDL script based on \citet{Horne86}.

Due to the curvature of the slit, it is inadvisable to combine the 2D spectra before the extraction and wavelength calibration step as is done in the current version of the CRIRES data reduction pipeline. 
The amount of curvature can be of the order of 1~pixel between two nodding positions at a nod throw of 10\arcsec and is variable in both spatial and spectral directions. We thus combined the individual 
1D spectra after correcting for the slit curvature by producing a telluric model spectrum using the method outlined in Sect.~\ref{sec:Synthesis} and interpolating the wavelength calibrated spectra to
 a common wavelength vector. All reduction steps were performed individually for the four chips of the CRIRES focal plane array.

\section{Spectral synthesis} \label{sec:Synthesis}
To compute theoretical transmission and radiance spectra of the Earth's atmosphere, a radiative transfer code is used that takes as
an input a model of the atmosphere in terms of vertical temperature, pressure and molecular abundance profiles, as well as line data 
(frequency, line strength, pressure broadening coefficients, etc.). In this section we describe the usage of such a code as well as the 
necessary inputs in further detail.

\subsection{Radiative transfer code}
There are several radiative transfer codes tailored for the construction of telluric spectra. These codes are based on line-by-line computations of a layered model of the Earth's atmosphere.

STRANSAC \citep{Scott74} is one the oldest of such codes. It utilises the GEISA spectral line catalogue \citep{GEISA99,GEISA03}. 

The 4A (Automatized Atmospheric Absorption Atlas) code also uses GEISA, and it provides a fast and accurate line-by-line radiative transfer model 
that is particularly efficient in the infrared region between 3,000\,nm and 17,000\,nm. The latest version of 4A can be retrieved free of charge for 
non-commercial usage from NOVELTIS\footnote{\texttt{http://www.noveltis.net/4AOP/}}, France. 

ATRAN \citep{Lord92} is another code to compute synthetic spectra of atmospheric transmission. It is based on a fixed atmospheric layering 
(see Sect.~\ref{sec:atmos}) and uses the HITRAN database (see Sect.~\ref{sec:database}) for line data input.

RFM (Reference Forward Model) is a line-by-line radiative transfer model originally developed at Oxford University\footnote{\texttt{http://www.atm.ox.ac.uk/RFM/}}, 
under an ESA contract to provide reference spectral calculations for the MIPAS instrument launched on the ENVISAT satellite in 2002. The code uses HITRAN 2000 as its line database.

In this paper we only consider FASCODE \citep{Clough81,Clough92}, and used both its commercial version FASCODE3P/PcLnWin\footnote{\texttt{http://www.ontar.com/Software/Products.aspx}}
and the free of charge version LBLRTM\footnote{\texttt{http://rtweb.aer.com/lblrtm\_description.html}}. LBLRTM is available as FORTRAN source code and can be 
compiled on various platforms. Its most common application is the spectral retrieval from ground and satellite based measurements in climatology. The code accepts a 
custom model atmosphere and uses the HITRAN database (see Sect.~\ref{sec:database}) as input for the line data. We found LBLRTM easy-to-use, yet flexible, and computed
all spectra presented in this paper with this code.

\subsection{Input I: Model atmosphere}\label{sec:atmos}
The choice of a representative model atmosphere is crucial for a realistic theoretical spectrum of the atmosphere.
One of the more basic descriptions of a model atmosphere is given by the 1976 US Standard Atmosphere\footnote{\texttt{http://ntrs.nasa.gov/archive/nasa/casi.ntrs.nasa.gov/\\19770009539\_1977009539.pdf}},
and its geographical subdivisions (tropical, mid-latitude and sub-arctic summer and winter models). Due to the
geographical location of Cerro Paranal and the fractional changes in the abundance of trace gases since its definition,
the 1976 US Standard Atmosphere gives a rather poor representation of the atmospheric conditions above Cerro 
Paranal.
\begin{figure}[!th]
\centering
\resizebox{\hsize}{!}{
\includegraphics[]{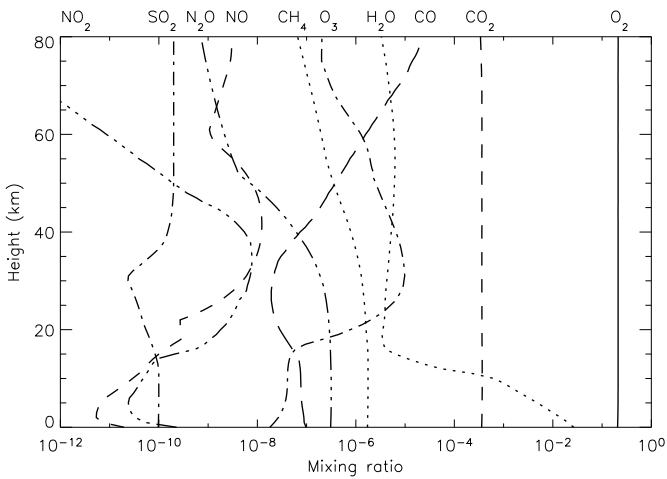}}
\caption{Vertical profiles of the 10 most abundant molecules included in our model atmosphere. The water vapour (H$_2$O) abundance 
for the lowest atmospheric layers (surface height $\leq$ 26\,km) is replaced by a meteorological model (GDAS or MM5) for the time of observation. 
\label{fig:equ_profile}}
\end{figure}
For example, the average surface temperature of the model at 2640\,m is only 271\,K and thus too low by 
about 10\,K on average.

In addition to the semi-static distribution of atmospheric trace gases, one has to account for the slowly variable 
temperature and pressure levels as well as the highly variable water vapour content. Thus, one has to base the 
calculations on a more flexible and realistic model atmosphere which takes meteorological inputs into account. 

We consider here the use of meteorological models from the Air Resources Laboratory (ARL) at the US National Oceanic 
and Atmospheric Administration (NOAA). Sounding files for temperature, pressure and dew point temperature in the troposphere and lower stratosphere (surface height $\leq$ 26\,km) 
are available at the ARL website\footnote{\texttt{http://www.arl.noaa.gov/ready/cmet.html}} based on either Global Forecast System (GFS) or archived Global Data 
Assimilation System (GDAS) models. GFS analysis data are available in 3~hr intervals for the next 84\,hours and 12~hr intervals for the next 384\,hours. GDAS models 
are available in 3~hr intervals for all dates since Dec 2004. 

The horizontal resolution of the GFS and GDAS models is 1\degr\, (approx. 110\,km). Given that Cerro Paranal is 
very close to the Pacific ocean ($\sim$12\,km), the influence of the ocean climate on the chosen grid point could 
compromise its validity. We have thus also computed model atmospheres for three nights during CRIRES SV periods on 
October 10, 2006, and March 3--4, 2007, based on the Fifth-Generation NCAR / Penn State Mesoscale Model (MM5) \citep{MM5}, which uses the
operational analyses of the European Centre for Medium-Range Weather Forecasts (ECMWF) as start and boundary conditions and was refined for the local 
topography of the Paranal area with a final resolution of 1\,km in the inner integration area. 
Both meteorological models reproduced the ambient pressure and temperature at the VLT as recorded in the file headers 
of the analysed observations to within 5\,hPa and 1.5\,K, respectively.

All atmospheric models have to be supplemented with information on the vertical distribution of 
all molecules other than H$_2$O and used in the radiative transfer calculations. Moreover, since the meteorological models are limited 
to estimated surface heights of $\sim$26\,km, we also have to add temperature, pressure and water vapour profiles 
for the remaining atmospheric layers ($26\leq h \leq 75$\,km). For this purpose we have chosen an equatorial MIPAS 
model atmosphere\footnote{\texttt{http://www-atm.physics.ox.ac.uk/RFM/atm/}}, 
constructed by John Remedios (U. Leicester). This model contains temperature, pressure and abundance information
for 30 molecular species in 121 levels from 0--120\,km at 1\,km spacing. In Fig.~\ref{fig:equ_profile} we show the 
mixing ratios of the 10 most abundant molecules in this model. The actual amount of molecules per cm$^3$ drops
roughly exponentially with the surface height and the predominant part of the telluric transmission is shaped for most species
in the first few km above the observatory.

\begin{figure}[!hb]
\centering
\resizebox{\hsize}{!}{
\includegraphics[bb=0 10 283 390]{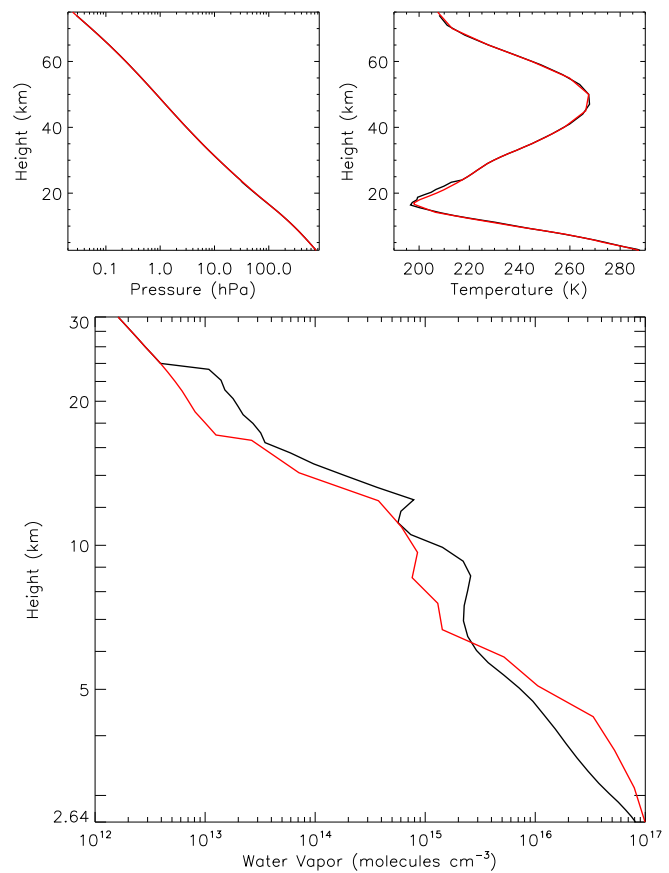}}
\caption{Comparison of atmospheric models for Cerro Paranal, GDAS (red) and MM5 (black). 
Shown are pressure (upper left panel), temperature (upper right panel) and water vapour (lower panel) profiles for 
March 3, 2007 at 09:00 UT for surface heights above Paranal (2.64\,km). 
Information for heights greater than 26\,km (GDAS) and 22\,km (MM5) were delivered by a global model for 
equatorial regions.
\label{fig:atmos}}
\end{figure}

In Fig.~\ref{fig:atmos} we compare our modified MM5 model to a GDAS model for March 3, 2007 at 09:00 UT. The MM5 was used with a set of modifications
developed by \citet{Zaengl02, Zaengl03} in order to improve the numerical accuracy over steep topography. While temperature and pressure profiles are nearly identical, the water vapour profile is noticeably different.
The GDAS model predicts a total water vapour column, PWV\footnote{measured as the precipitable water vapour 
amount at zenith}, of 4.1\,mm, the modified MM5 model predicts 2.4\,mm. Moreover, the vertical distribution of the water vapour is different as well. The GDAS 
model predicts a much more humid surface layer, but dries out beyond the prediction of the MM5 model above 6\,km height. A similar situation was found for October 
10, 2006. Both models predict the same PWV of 1.8\,mm but the GDAS model is again more humid below 3\,km and less humid above. In any case, the models are 
often too wet when compared to the observations. The total water vapour amount was always scaled by 40\%--70\% to match the observed line depth (see \S4). 
The different vertical distributions of the water vapor did not lead to a significant difference in the fit quality of the water vapor lines once the total water vapor amount 
was rescaled to reach the best fit. We compare the predicted and fitted PWV values in Tab.~\ref{tab:h2o} and conclude here, that the GDAS models are applicable and 
sufficient as meteorological input to refine the global equatorial model atmosphere. See Sect.~\ref{sec:performance} for a further discussion about the performance and 
limitation of the actual fitting.

\subsection{Input II: Molecular line database}\label{sec:database}
HITRAN, in its latest 2008 edition \citep{HITRAN2008}, is a molecular line database containing frequency,
line strength and pressure broadening coefficients for more than 1.7 million spectral lines of 42 different molecules
and their common isotopes. The database is regularly updated and supplemented with new line data and molecular 
cross-sections. LBLRTM is distributed with a line file that is based on the HITRAN 2004 version and includes individual 
updates for some molecules until June 2009. A complete update to HITRAN 2008 is planned for the near future 
(Mark Shephard, private communication). 

\section{Results}\label{sec:results}\label{sec:performance}
We have computed synthetic telluric transmission spectra for a number of selected CRIRES observations 
(see Tab.~\ref{tab:obslog}) using LBLRTM and a GDAS model atmosphere closest in time to the observation. 
The wavelength range of the model was set to the coverage of the chosen CRIRES setting. LBLRTM uses the 
altitude angle of the telescope pointing, the height of the observatory and the model atmosphere to 
calculate the line-by-line and layer-by-layer transmission and radiance. We make no use of the calculated radiance
at this point, since the telluric emission is removed conveniently by nodded observations, and use the transmission
spectrum only.

An IDL based Marquardt--type $\chi^2$ minimisation code \citep[][p. 161]{Bevington} was used to (a) refine 
the wavelength solution by fitting a second order polynomial, using the original wavelength calibration as 
first guess; (b) re-determine the total water vapour amount and column density of other molecules present 
in the chosen setting; and (c) determine the instrumental profile and continuum rectification parameters. 

\begin{figure}[!hb]
\centering
\resizebox{\hsize}{!}{
\includegraphics[]{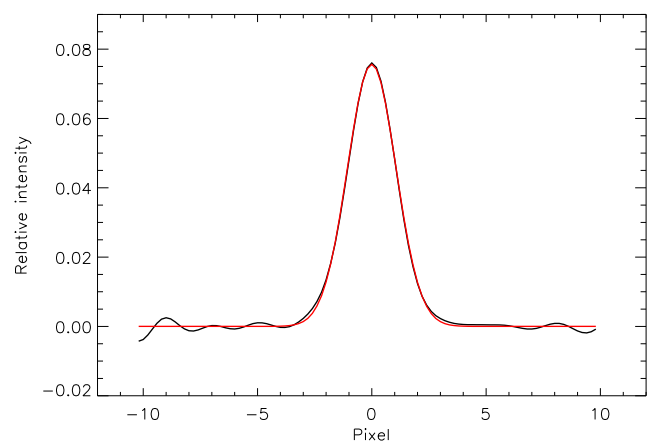}}
\caption{CRIRES instrumental profile, obtained from SVD fitting of methane (CH$_4$) lines in setting \lref 2336\,nm
(see Fig.~\ref{fig:2336}). Black: empirical profile constructed from 21 Eigenvalues. Red: best fit Gaussian model with a FWHM
of 2.67\,pixel (4.0\,km/s, $R$=70,000).}
\label{fig:IP}
\end{figure}

The determination of the instrumental profile (IP) is crucial for a proper fitting of the only partially resolved
atmospheric lines. We have used a singular-value decomposition (SVD) technique in an implementation given
by \citet{SVD} to determine the instrumental profile. The wavelength scale was interpolated to a five times 
higher sampling than the original, Nyquist sampled wavelength scale and a 20\,pixel wide instrumental profile
 was constructed from 11--13 Eigenvalues from all telluric lines present in the spectrum. 
Higher number of Eigenvalues did not lead to a further improvement of the fit but amplified the noise in the instrumental
profile. In all cases, the instrumental profile was extremely well matched by a single Gaussian function
whose FWHM depends on the the slitwidth and the performance of the adaptive optics, see Fig.~\ref{fig:IP}. 

In cases where extremely narrow spectral lines were present (e.g. the unresolved ozone lines at 4700\,nm - 
see Sect.~\ref{sec:4760}) up to 30 Eigenvalues could be used without amplifying the noise.
This analysis confirmed that a Gaussian model is a valid approximation for the instrumental profile at signal-to-noise levels below $\sim300$. 
We have thus used a single Gaussian model for the final fitting of the telluric model, the width of the profile being the 
only free parameter. We note that our determined FWHM are broader than expected for $R=100\,000$. At the time most 
of the data presented in this paper were recorded, the focal plane array of CRIRES was slightly out of focus. Thus, 
the maximum resolving power measured in the presented dataset is $R\sim87\,000$.

\subsection{Performance in selected settings}
We present CRIRES observations of bright standard stars that contain clear and isolated line systems of the most abundant molecules of the Earth's atmosphere. 
We evaluate the performance of the telluric correction by checking the residuals for systematics by a rigid quantisation of the signal-to-noise ratio (SNR) in the 
corrected spectra. We chose the SNR as a measure for the performance of the telluric model fit since it has the strongest practical implication for the observer,
as it allows a judgement as to which precision a position or an equivalent width of a stellar line can be measured in a spectrum. In regions of telluric contamination, 
the residuals from the removal of telluric lines are the main factor hindering a high signal-to-noise measurement, otherwise limited only by the source flux or the 
uniformity of the detector response.
   
In order to compare the SNR before and after the correction, we first compute the intrinsic noise level of our CRIRES spectra by adding the photon noise of the 
source and the measured background noise in quadrature for each pixel along the dispersion direction. The background noise consists of the detector readout noise, 
the photon noise of the dark current and the photon noise of the thermal emission of the atmosphere and the telescope after pairwise subtraction of nodded frames. 
We have validated this procedure by measuring the standard deviation in the continuum level of each spectrum. The result was always consistent with our noise 
estimate. 

Hence, for the input spectra we have the SNR as a function of wavelength which scales with the square root of the received flux, and thus with the square root 
of the atmospheric transmission $T$.  In addition, the telluric emission lines in the thermal infrared example ($\lambda\sim4700$\,nm, Sec.~\ref{sec:4760}) add 
noise where the lower transmission of the same telluric lines in absorption degrade the SNR in comparison to the continuum level. 

The fitting procedure minimises the residuals in the observed$-$computed spectrum (O$-$C) over the whole wavelength region of the considered spectral 
range of $\sim\lambda/200$, recorded on one CRIRES detector. For the removal of the telluric lines, however, the measured spectrum has to be divided by the 
computed model. After division by the model, we measure the noise in the spectrum as the standard deviation of all pixels. The SNR is then simply the inverse of 
the measured noise, given that after the division the signal level is at unity for a normalized spectrum. 

It is important to remember that even for a perfect model of the telluric lines the O/C residuals will be larger, and the SNR smaller, in comparison to the undisturbed 
continuum, since the spectrograph receives less flux where telluric absorption lines are present. Hence, when dividing through the telluric transmission spectrum, and 
thus correcting the received flux for the presence of these lines, the noise in these regions is higher than in the neighboring continuum, simply because less flux was 
received in the lines in the first place. Albeit this caveat, we use the formal SNR as a direct measure of the performance of the modeling since it reflects the impact 
of the telluric lines on the astronomical spectrum and gives the observer a meaningful number to work with.

\subsubsection{H$_2$O at 1505\,nm}\label{sec:1505}
\begin{figure}[!ht]
\centering
\resizebox{\hsize}{!}{\includegraphics[]{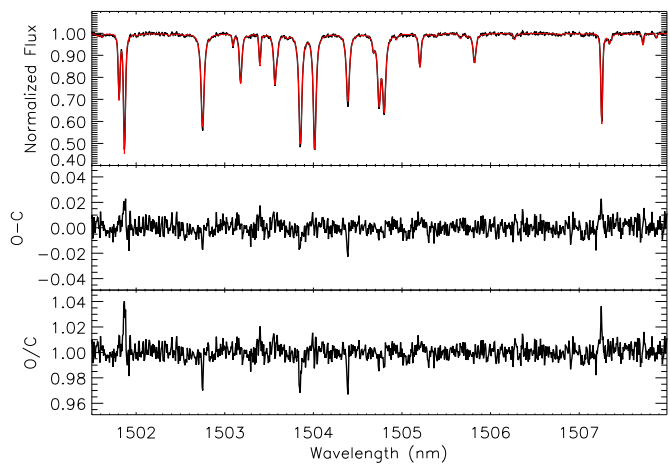}}
\caption{Water vapour (H$_2$O) lines on chip three of CRIRES setting \lref 1504.8\,nm. 
Top panel: CRIRES spectrum (black), and synthetic transmission spectrum (red). 
Lower panels: O-C and O/C residuals of the fit.
\label{fig:1504}}
\end{figure}

Water vapour (H$_2$O) is the most important absorber in the near-infrared. In addition to the strong $\nu_1$ and $\nu_3$ fundamental bands at 2700\,nm, 
several overtone and combination bands centered at 940\,nm, 1100\,nm, 1380\,nm, 1870\,nm, and 3200\,nm are present in near-infrared spectra. We picked 
a few isolated, unsaturated H$_2$O lines at 1505\,nm in a CRIRES standard star setting from March 03, 2007. The airmass of the observation was 1.033, 
the resolving power was $\sim$80,000 at a slitwidth of 0.2\arcsec. Both the GDAS model and the modified MM5 model had to be scaled to a PWV of 1.6\,mm 
for the best fit with the observation (see Fig.~\ref{fig:1504}).
The achieved correction using the GDAS and MM5 models is generally satisfying but falls behind the expectation from the noise limit. The average SNR of the measured 
spectrum is $\sim$200. After the division by the model spectrum, the average SNR drops to $\sim$180. In the line centers (40-70\% transmission) the SNR drops from 
$\sim$160 to $\sim$50 after dividing by the model. Note how lines of equal strength, e.g., around 1504~nm are fitted with greatly varying precision, pointing to 
uncertainties in the line parameters.

\subsubsection{CO$_2$ at 2060\,nm}\label{CO2}
Carbon dioxide (CO$_2$) has a nearly constant vertical mixing ratio (see Fig.~\ref{fig:equ_profile}) but due to its strong interaction with the biosphere, the low altitude
CO$_2$ abundance can be expected to be variable in time. CO$_2$ is the third most abundant molecule after oxygen and nitrogen at heights exceeding $\sim$4\,km 
and exhibits various strong absorption bands centered at 1400\,nm, 1600\,nm, 2000\,nm, 2700\,nm, and 4300\,nm. \citet{Kenworthy} explored the performance of 
empirical telluric line removal around the important \ion{He}{I} lines at 2058.69\,nm, which is buried under a comb of lines from the $R$ branch of the 
20013$\leftarrow$00001 transition of telluric \element[][12]{C}\element[][16]{O}\element[][16]{O}. 

\begin{figure}[!ht]
\centering
\resizebox{\hsize}{!}{
\includegraphics[]{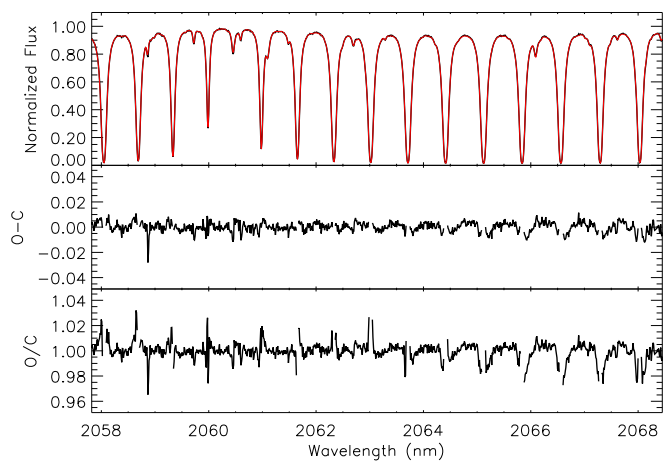}}
\caption{Same as Fig.~\ref{fig:1504} but for carbon dioxide (CO$_2$) lines on chip two of CRIRES setting \lref 2076.5\,nm. 
Regions with telluric transmission of less than 5\% are not shown in the residuals.
\label{fig:2076}}
\end{figure}

We have used a standard star observation obtained on March 01, 2007 in this spectral region to test the performance of our synthetic telluric model. The airmass of the 
observation was 1.24, the resolving power was $\sim$60,000 at a slitwidth of 0.4\arcsec. The CO$_2$ abundance is given by the global atmospheric model (see Sect.~\ref{sec:atmos}) 
and had to be scaled down by $\sim$7\% to get the best match with the observation (see Fig.~\ref{fig:2076}), despite the fact that the atmospheric CO$_2$ abundance is 
continuously growing and shows yearly changes of only $\pm$1\% (NOAA/ESRL\footnote{\texttt{www.esrl.noaa.gov/gmd/ccgg/trends/}}).

The average SNR of the spectrum is $\sim$300. It drops to about 230 at transmissions $\geq$ 60\% when corrected for the telluric absorption. The SNR at the line cores 
(10--50\% transmission) is $\sim$160 and drops to $\sim$50 after the correction, see Fig.~\ref{fig:2076}. As already noted by \citet{Kenworthy}, a precise wavelength 
solution is crucial for a good fitting. Wavelength errors at the level of 1/50\,pixel ($\sim$30\,m/s) notably degrade the fitting quality because of the steep line profiles. 
While the line positions in the HITRAN database have uncertainties at the m/s level, the uncertainty in the pressure shifts may become an issue at this level of precision. 

After dividing by the best fit synthetic model, the region around the \ion{He}{I} line has a SNR of $\gtrsim$160. We note that a high-$J$ line of \element[][13]{C}O$_2$ 
leaves a 3\% residual right at the rest wavelength of the \ion{He}{I} line.  We also note that for increasing $J$ quantum numbers, the line shape of the \element[][12]{C}O$_2$ 
is less well matched as close to the band center. CO$_2$ exhibits strong line coupling which is taken into account by LBLRTM (see discussion in Sect.~\ref{sec:Summary}). 
It remains thus unclear whether the systematic increase of the residuals are due to uncertainties in HITRAN or due to insufficient
treatment of the line coupling.

The \element[][12]{C}O$_2$ lines are fully resolved and only a few narrow lines are present in the setting. This inhibits a precise determination of the instrumental profile, 
which becomes very sensitive to noise in the spectrum. Thus, the width of the instrumental profile is partly degenerate with the line strength of CO$_2$ in the model. The exact 
locus of the best correction can be modified by slight changes of the FWHM of the instrumental profile. This favours a customised telluric correction, tailored for a specific stellar line.

\subsubsection{CH$_4$ at 2320\,nm}
Methane (CH$_4$) exhibits strong bands centered at 1660\,nm, 2200\,nm, and between 2300\,nm to 3830\,nm, although it is much less abundant than H$_2$O and CO$_2$. 
The methane abundance in our model is set by the global atmospheric model (see Sect.~\ref{sec:atmos}). Methane is another of the so called \textit{greenhouse gases}. Its 
atmospheric concentration is slowly increasing with time, but its growth rate dropped considerably in the last years \citep{methane}. 
\begin{figure}[!ht]
\centering
\resizebox{\hsize}{!}{
\includegraphics[]{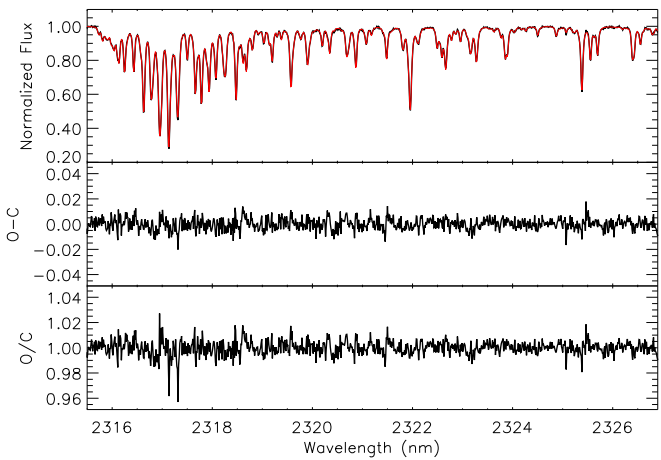}}
\caption{Same as Fig.~\ref{fig:1504} but for methane (CH$_4$) and a few water vapour (H$_2$O) lines on chip two of CRIRES setting \lref 2336.2\,nm. 
\label{fig:2336}}
\end{figure}
We chose a observation from Feb 29, 2008, covering a strong CH$_4$ band at 2320\,nm for fitting. The airmass of the observation was 1.01, the resolving power was $\sim$65,000 
at a slitwidth of 0.4\arcsec. This region is populated by CO first overtone lines in astrophysical objects and thus marks a very interesting spectral region. Besides the strong CH$_4$ lines, 
there are also three prominent water vapour lines in the covered region, one strong line at 2321.95\,nm, and two weaker line blends at 2322.66\,nm and 2323.27\,nm. The best fit PWV 
at zenith was 2.2\,mm. We did not have to scale the original CH$_4$ abundance in the equatorial model. The average SNR of the spectrum is $\sim$300, dropping to $\sim$220 when 
corrected for the telluric absorption.

\subsubsection{N$_2$O at 3907\,nm}
An observation from March 05, 2007 covered a nitrous oxide (N$_2$O) band around 3907\,nm at airmass 1.3. The resolving power was only $\sim$65,000 at a slitwidth of 0.2\arcsec.
N$_2$O is a factor of two less abundant than CH$_4$. Still, it exhibits strong line systems between 2870 and 4500\,nm. As for CO$_2$, we had to scale the N$_2$O abundance in our 
model down by $\sim$5\% to match observations. In addition to the strong N$_2$O lines, there are weaker lines of CH$_4$ and H$_2$O present in this setting. The best fit PWV at 
zenith was 8.6\,mm. The average SNR of the spectrum is $\sim$380, dropping only by 20\% to $\sim$310 when corrected for the telluric absorption. 
\begin{figure}[!hb]
\centering
\resizebox{\hsize}{!}{
\includegraphics[]{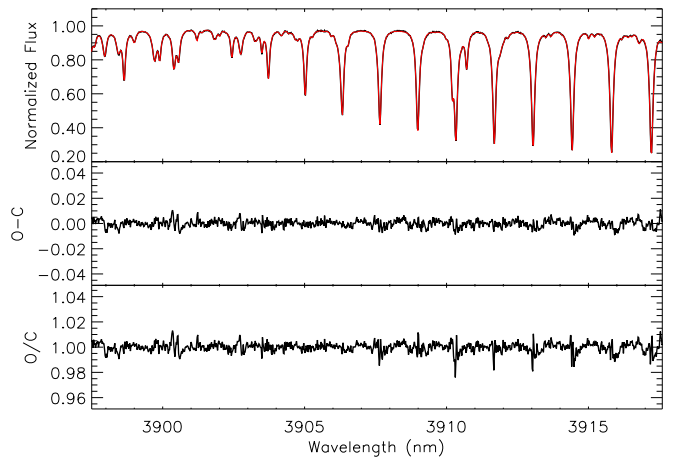}}
\caption{Same as Fig.~\ref{fig:1504} but for nitrous oxide (N$_2$O) lines on chip three of CRIRES setting \lref 3907.2\,nm. Minor contributions between $\lambda\lambda$ 3898--3905\,nm 
arise from H$_2$O and CH$_4$. 
\label{fig:3907}}
\end{figure}

\subsubsection{H$_2$O, O$_3$, CO$_2$, and CO at 4700\,nm}\label{sec:4760}
\begin{figure}[!hb]
\centering
\resizebox{\hsize}{!}{
\includegraphics[]{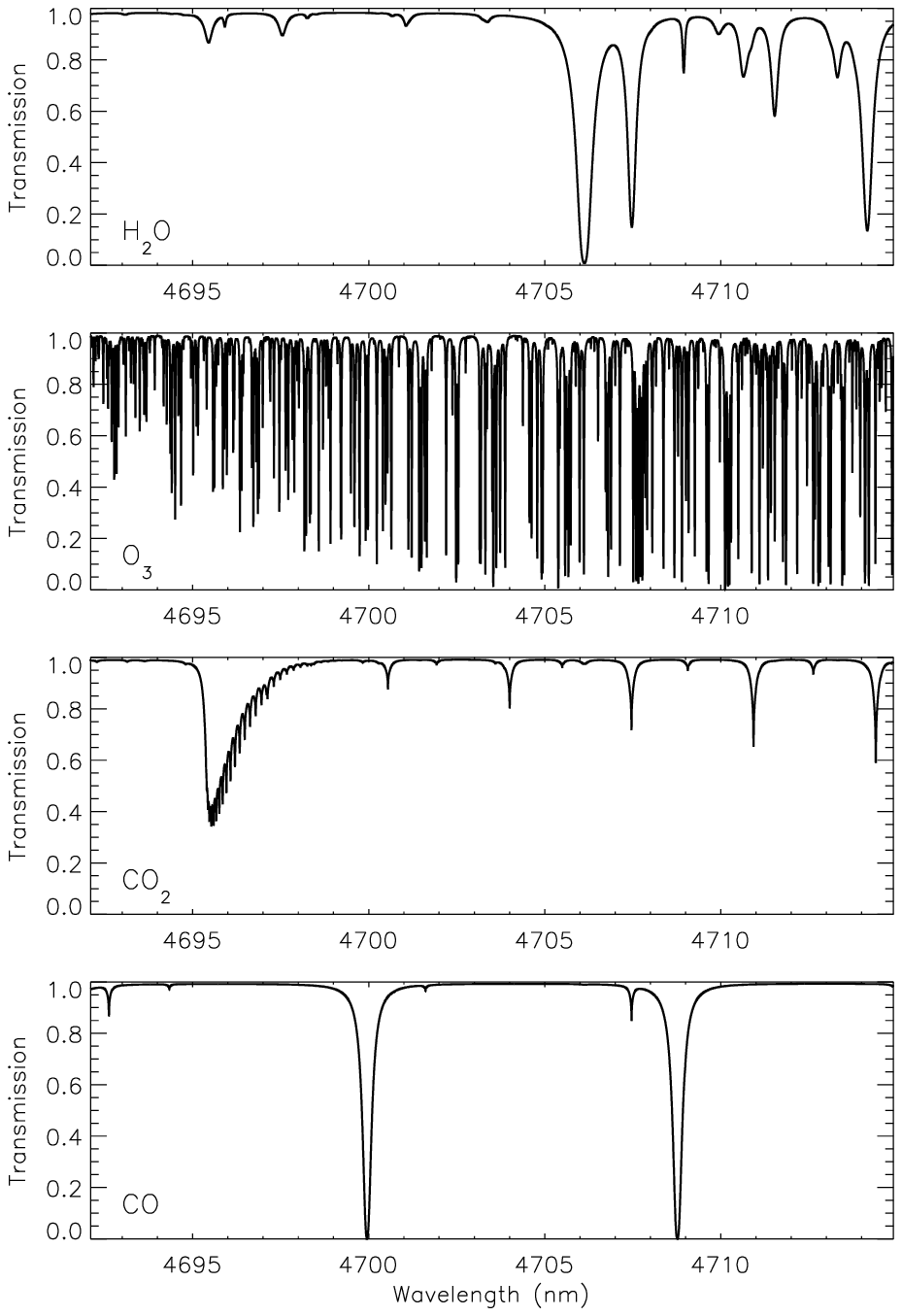}}
\caption{Transmission of the individual species involved in the spectral region shown in Fig.~\ref{fig:4760}. From top
to bottom: water vapour (H$_2$O), ozone (O$_3$), carbon dioxide (CO$_2$), and carbone monoxide (CO). The transmissions are
shown before convolution with the instrumental profile. 
\label{fig:4760_species}}
\vspace{0.6cm}
\resizebox{\hsize}{!}{
\includegraphics[]{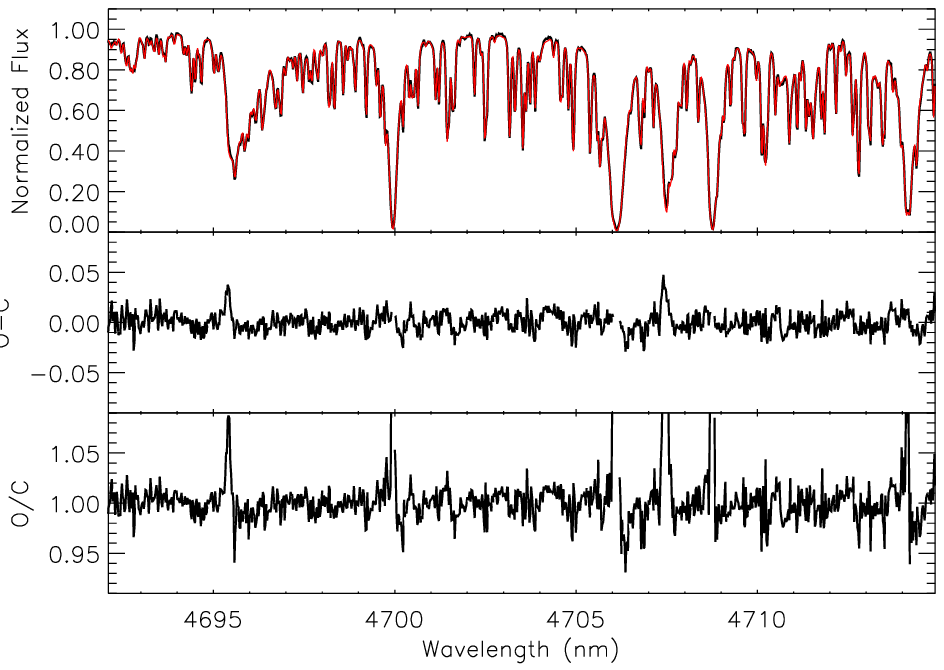}}
\caption{Same as Fig.~\ref{fig:1504} but for ozone (O$_3$), water vapour (H$_2$O), carbon dioxide (CO$_2$) 
and carbone monoxide (CO) on chip one of CRIRES setting \lref 4760.8\,nm. 
\label{fig:4760}}
\end{figure}

Ozone (O$_3$) is a major opacity source in the UV where the Hartley and Huggins bands block most of the stellar and solar radiation. The Chappuis bands at 600\,nm are weak, but still visible 
in optical spectra \citep{Griffin05} and originate from the stratosphere where ozone has its maximum abundance at about 30\,km height (see Fig.~\ref{fig:equ_profile}). The near-infrared band of 
O$_3$, centered at 4750\,nm, is produced by overtone and combination transitions. These lines are strong but comparably narrow, which is due to the low atmospheric pressure at the height of
 formation. 

We chose a standard star observation from April 22, 2007 covering a region around 4700\,nm at airmass 1.06. The resolving power was $\sim$87,000 at a slitwidth of 0.2\arcsec. In the chosen 
region, H$_2$O, CO$_2$, and CO lines contribute to the opacity, see Fig.~\ref{fig:4760_species}. The $Q$ branch of the 20001$\leftarrow$01101 transition of CO$_2$ is very prominent 
(see the second to last panel in Fig.~\ref{fig:4760_species}). 

The fundamental transitions of carbon monoxide (CO) are of great importance for astronomical observations. Unfortunately, the CO lines are also very strong in the telluric spectrum and the transitions 
with $2 \leq J \leq 12$ are saturated, even at modest airmasses. High spectral resolution and velocity offsets from barycentric motion and radial velocity of the science target are needed to disentangle the stellar and telluric components. 
The average SNR of the measured spectrum in Fig.\ref{fig:4760} is $\sim$200, which dropped to $\sim$100 when divided by the model spectrum. The $CO_2$ $Q$ branch 
is not well fitted and peaks out as a strong residual near 4695~nm. The second strongest residual near 4707.5~nm is a H$_2$O line. This line was much better fitted with the 
line data from the HITRAN 2004 database before the update in 10/2008 (AER V2.1).  Its line strength is notably different in the newest line file (AER V2.2) and causes a worse 
fit. Other minor residuals are mostly from O$_3$. Because the line formation happens in a cold part of the Earth's atmosphere at $\sim$20--30\,km height and its abundance sharply 
decreases below 15\,km, O$_3$ does not contribute notably to the telluric emission received at the ground but is a strong source of absorption. Only small scaling corrections at the 
order of 2.5\%--3\% had to be applied to the model abundances of CO$_2$, and CO while the O$_3$ abundance had to be increased by 6\%. The best fit PWV was 7.2\,mm (7.7\,mm 
using the HITRAN 2004 database).

\subsection{Limitations}
\begin{figure}[!ht]
\centering
\resizebox{\hsize}{!}{
\includegraphics[]{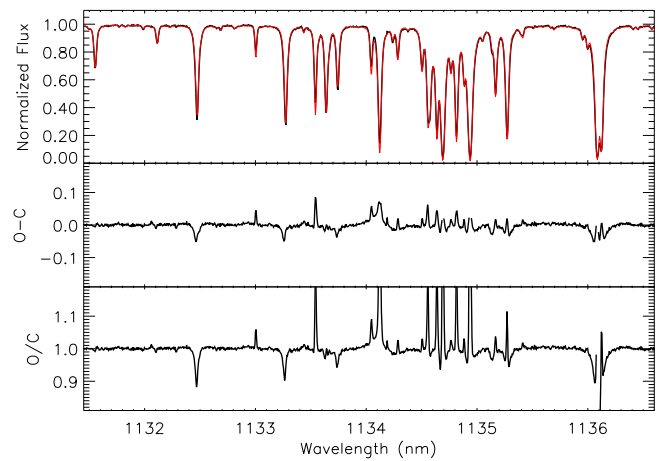}}
\caption{Same as Fig.~\ref{fig:1504} but for water vapour (H$_2$O) lines on chip three of CRIRES setting \lref 1134.0\,nm. 
\label{fig:1134}}
\end{figure}

For the cases presented in the previous subsection, we have never reached the formal noise limit of the observations but we found a satisfying fit of the model to the data. 
To illustrate the limits of our approach, we present an extreme example of very badly fitting H$_2$O lines in Fig.~\ref{fig:1134}. The airmass of the observation was 1.3, 
the resolving power was $\sim$64,000 at a slitwidth of 0.2\arcsec. This is the worst fit we have come across and it is clearly related with the line strength information in 
HITRAN. Neighboring lines in the model are both too strong and too weak. The predicted PWV at zenith was $\sim$2\,mm and was scaled down during the optimisation to 
0.35~mm. We find this reasonable, since the PWV was reported at the same level for other spectral settings taken during the same night where the water vapour lines 
were fitted much better. The lines in the setting shown in Fig.~\ref{fig:1134} are thus very strong lines as they show up with an average transmission of 50\% and below 
even under such dry conditions. Water vapour lines also caused problems in the setting presented in Sec.~\ref{sec:4760}, and left residuals in other settings. It is thus the 
water vapour which appears the hardest to model and to be affected the most from uncertainties in the line database. 

\subsubsection{Limitations from the HITRAN database}
The HITRAN database contains estimators on the uncertainty of position, line strength and pressure broadening for all entries \citep{HITRAN2004}. For example, the
\element[][12]{C}O$_2$ lines shown in Fig.~\ref{fig:2076} have an average uncertainty in their positions of $\sim5\times10^{-5}$\,cm$^{-1}$ \citep{Miller04} or 
$\sim3$~m/s and the uncertainty in the line strength is at the few-\% level \citep{Tashkun03}. The strongest CH$_4$ lines shown in Fig.~\ref{fig:2336} have an 
average uncertainty in their positions of $\sim10^{-3}$\,cm$^{-1}$ or $\sim60$~m/s and uncertainties in their line strengths of up to 10\% \citep{HITRAN2004,Brown2003}. 
This level of positional uncertainties influences the wavelength solution and the line fitting. As expected, even the strongest water vapour lines shown in Fig.~\ref{fig:1134} 
have positional uncertainties of up to 0.01\,cm$^{-1}$ or $\sim2.6$~km/s and their line strength can be uncertain to more than 20\% \citep{HITRAN2004}. This explains 
the bad fit to the data shown in Fig.~\ref{fig:1134}. 

\subsubsection{Limitations from LBLRTM}
Other limitations arise from the limited treatment of line mixing (aka line coupling) effects and from the assumed line profile. Line mixing in LBLRTM is modeled using a first 
order perturbation approach. For CO$_2$, line mixing is treated according to \citet{Niro05}. LBLRTM employs a Voigt line shape at all atmospheric levels with an algorithm 
based on a linear combination of approximating functions \citep{Clough05}. A Rautian, Galatry or speed-dependent Voigt profile would be needed to account more precisely 
for speed-dependent broadening effects and line mixing effects \citep[e.g.,][Fig.1]{Brault03}. The errors induced by this simplification are, however, comparatively small 
against the uncertainties in other line parameters entries of the HITRAN database.

\subsubsection{Limitations from the model atmosphere}
As long as the total column densities of the different molecular species can be varied, the uncertainties in the model atmosphere have only a minor impact on the model 
spectrum. Exceptions are CO and O$_3$, where changes in the vertical abundance profiles lead to notably different line widths, because either the abundance maximum or 
minimum of the species is close to the temperature inversion of the Earths atmosphere (see Fig.~\ref{fig:equ_profile} and \ref{fig:atmos}). The model for our mid-infrared 
case presented in Sec.~\ref{sec:4760} does show a dependence on the type of model atmosphere, i.e. whether an equatorial or night-time MIPAS model is used.  The 
intrinsic line width is, however, degenerate with the instrumental profile and while the formal resolution measured as the width of the instrumental profile varies, the residuals 
give no hint which of the two model atmospheres is more representative of the actual conditions. 

For the water vapour, we did not find a significant improvement in the fit quality when using the MM5 atmospheric models instead of the GDAS models. The fitting 
delivered the same PWV for both models and the different vertical distribution of the water vapour had no noticeable influence on our test cases. Re-fitting of the vertical 
distribution of the water vapour will thus lead only to minor improvements, if at all, and seems only feasible when enough lines are present, preferably from different vibrational 
transitions. 

\subsubsection{Limitations from stellar lines}
The fitting of a telluric model to an observed spectrum will be more complicated in the presence of blends between telluric and stellar lines. Various strategies could be used 
to allow for the same fitting quality as compared to the featureless early type stars used in this study. (I) The usage of a stellar model spectrum in the fitting process would 
significantly improve the telluric model. (II) In the thermal infrared, the telluric emission lines could be used to further constrain the fit in terms of wavelength solution and 
molecular abundances. (III) In the worst case, an observation of a telluric standard star in the same spectral setting but not necessarily at the same time and airmass can be 
used to constrain the model fit for the target \citep[e.g.,][]{Villanueva}.

\section{Conclusions}\label{sec:Summary}
  \begin{table*}[!ht]
      \caption[]{Precipitable water vapour values (in mm PWV at zenith).}\label{tab:h2o}
\centering 
\begin{tabular}{c r l c c c} 
\hline\hline 
Date \& Time & \multicolumn{2}{c}{GDAS} & \multicolumn{2}{c}{MM5} & ESO$^\mathrm{a}$ \\ 
     (UT)    &predicted&fitted&predicted&fitted&predicted\\
\hline 
2006-10-10 03:00&  1.8 & 0.8  & 1.8  & 0.8  & 1.9 \\ 
2007-03-01 06:00&  5.3 & 2.6  & -    & -    & 3.1 \\
2007-03-03 09:00&  4.1 & 1.6  & 2.4  & 1.6  & 1.2 \\
2007-03-05 06:00& 10.8 & 8.6  & -    & -    & 3.9 \\
2007-04-22 09:00&  9.6 & 7.2  & -    & -    & 2.5 \\
2007-05-10 01:00&  2.0 & 0.4$^\mathrm{b}$  & -    & -    & 3.0 \\ 
2008-02-29 09:00&  3.6 & 2.2  & -    & -    & 2.8\\
\hline 
\end{tabular}
\begin{list}{}{}
\item[$^{\mathrm{a}}$] Prediction for Paranal Observatory, based on satellite measurements. \\ 
Obtained from \texttt{http://www.eso.org/gen-fac/pubs/astclim/forecast/meteo/ERASMUS/} 
\item[$^{\mathrm{b}}$] PWV determined from water vapour lines not shown in Fig.~\ref{fig:1134}.
\end{list}
\end{table*}
We have shown that with the use of LBLRTM, a line-by-line radiative transfer code, and an appropriate model atmosphere, synthetic telluric transmission spectra 
can be constructed that match the observed telluric spectrum to 2\% or better (see, e.g., Fig.\ref{fig:2076}), even close to saturated lines. These results 
were achieved with only a modest effort in back-fitting of atmospheric properties. Compilation of LBLRTM is straight forward and when using the described IDL fitting 
routines for $\chi^2$ minimisation, the time needed for a model fit to a single CRIRES chip is less than 2~minutes on standard desktop PC . User interaction is only 
required to download the appropriate GDAS model from the NOAA website, to set up an initial wavelength scale, and to choose the species to be fitted. 

Although full photon noise limited performance is never achieved, the usage of the synthetic transmission spectra offers an alternative to the usage of standard star observations. 
The tight constraints in matching airmass and instrumental profile of the science target imposed on telluric standard stars are hard to meet in reality without downgrading the 
achievable SNR or the spectral resolution. 

\begin{figure}[!hb]
\centering
\resizebox{\hsize}{!}{
\includegraphics[]{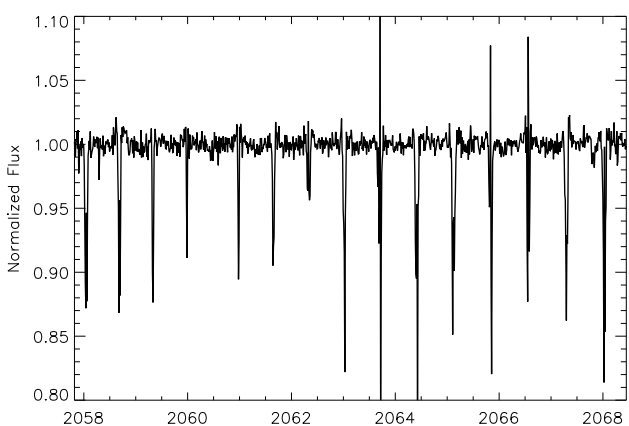}}
\caption{Division of two spectra from consecutive nodding positions of the same star shown in Fig.~\ref{fig:2076}. 
The spectrum from nodding position B was mapped on the wavelength scale of of the spectrum from nodding position A 
before the division. The difference in airmass is only 0.005. See Fig.~\ref{fig:2076} and text for details.
\label{fig:ab}}
\end{figure}

We demonstrate this problem by using the data shown in Sect.~\ref{CO2}, this time showing in Fig.\ref{fig:ab} the spectrum 
from nodding position A divided by the one in nodding position B, taken 1.5\,min later.
The difference in airmass is only 0.005. The spectrum of the B position had to be 
mapped onto the wavelength scale of the A position due to the slit curvature which changes the wavelength zero point and the dispersion between both positions. We note that 
at this step the wavelength solution provided by the telluric model for both nodding positions had to be used to achieve this result. Despite the fact that both spectra are of 
the same star and taken close in time and airmass, a slight change in the AO performance over time and in the instrumental profile between both nodding positions led to
the apparent residuals in the division. The detector response at the two nodding positions is uniform to better than 0.5\% after correcting for non-linearity 
and applying a flatfield correction as outlined in Sec.~2. Hence, changes in the instrumental profile are the reason for the observed residuals. 

Under more realistic circumstances, i.e. when a different target is
observed after a longer time interval and not matching the airmass that well, the expected residuals from a simple division by another telluric standard star would be even larger. 
Even in this very favorable case, the residuals in the line cores are at the order of 10\% and the residuals in the wings are comparable to the ones after dividing through a 
telluric model (see Fig.~\ref{fig:2076}). Acquiring accurate empirical telluric spectra with CRIRES is challenging and the usage of telluric model spectra is an interesting 
alternative to the observation of standard stars, especially since it also saves valuable observing time.

An effective removal of telluric absorption lines demands, but in most cases also provides, a wavelength calibration at an extremely high level of precision. Narrow and steep 
absorption lines are sensitive to wavelength errors at the level of 1/50~pixel (30~m/s). Having the telluric absorption lines directly imprinted in the science spectrum, allows
for the most unbiased wavelength calibration when compared to telluric emission lines or calibration lamp lines since the telluric absorption lines experience the same IP as 
the stellar lines. Given the large variation in accuracy of the line positions in the HITRAN database, a wavelength solution based on a weighted fit to telluric lines offers further 
improvements, especially when a sufficient number of lines is present. With the excellent line positions of the \element[][12]{C}O$_2$ lines shown in Fig.~\ref{fig:2076} a 
wavelength solution with an r.m.s. of $\leq20$\,m/s can be achieved. 

To a certain extent, the Earth's atmosphere serves as a 'gascell' for precise wavelength calibration as shown by \citet{Seifahrt08} who used N$_2$O lines at 4100\,nm and
reached a short term wavelength precision of $\leq10$\,m/s over several hours. 

Recently \citet{Figueira10} have demonstrated, based on 6 years of HARPS optical
data, that atmospheric O$_2$ absorption lines are stable down to 10\,m/s and can be used 
as a frequency standard with an internal precision of better than 2\,m/s. Transferring this
precision into the infrared wavelength regime needs a precise model for
the telluric interference spectrum to account for variability, mostly
H$_2$O content and airmass. The latter is not trivial in the presence of
fully or nearly fully saturated lines.

The technique described in the current paper was used to calculate telluric model spectra as part of the data analysis 
of a near Infrared radial velocity program conducted by \citet{Bean10a,Bean10b} with the CRIRES spectrograph. Used 
in conjunction with an Ammonia gascell, an RV precision of 3-5 m/s was obtained on mid-M dwarfs in a spectral region
near 2300\,nm.

Despite not being intended as a study for time dependent water vapour distribution, we find that the observed variability in the PWV above Paranal, ranging from 0.4\,mm to 
nearly 9\,mm in our observations, exceeds the range of satellite based predictions for the PWV by far. We have compiled all predicted and fitted PWV values used in this study 
in Tab.~\ref{tab:h2o}. The PWV can change notably in the course of hours and re-fitting of the total PWV for a specific observation is unavoidable. A detailed study on the 
water vapour column density above Paranal will be presented elsewhere (Smette et al., in preparation). The GDAS meteorological model very often predicts too high PWV values, 
which we have always scaled globally, without changing its vertical distribution. Comparing GDAS and MM5 models for the same nights, results in the same PWV value and the 
use of the more realistic MM5 model did not significantly improve or fit results. 

Choosing well characterised lines, we do see the potential in the CRIRES data to retrieve atmospheric abundance information for atmospheric trace gases at or below the 1\% level 
delivering important information about the short and long term variations of greenhouse gases from astronomical observations. Such measurements could be a valuable byproduct 
of astronomical spectra collected with high-resolution infrared spectrographs.

\begin{acknowledgements}
We thank all colleagues involved in the development, installation and commissioning of CRIRES. Special thanks goes to Alain Smette for his comments and suggestions
on the telluric modeling and Linda Brown for valuable discussions on line broadening and -mixing. A.S. acknowledges financial support from the Deutsche Forschungsgemeinschaft 
under DFG RE 1664/4-1. M.J.R. and A.S. acknowledge support from the National Science Foundation under grant NSF AST-0708074. J.L.B. acknowledges research funding from the
European Commissions Seventh Framework Programme as an International Incoming Fellow (grant no. PIFFGA-2009-234866).
\end{acknowledgements}

\end{document}